# Strong Long-Wave Infrared Optical Response in a Topological Semiconductor with a Mexican Hat Band Structure

Mark J. Polking,[1,*] Haowei Xu,[2] Raman Sankar,[3] Kevin Grossklaus,[4] and Ju Li[2,5,*]

[1] Advanced Materials and Microsystems Group, Massachusetts Institute of Technology Lincoln Laboratory, 02421, United States

[2] Department of Nuclear Science and Engineering, Massachusetts Institute of Technology, 02139, United States

[3] Institute of Physics, Academia Sinica, 11529 Taiwan

[4] Department of Electrical and Computer Engineering, Tufts University, 02155, United States

[5] Department of Materials Science and Engineering, Massachusetts Institute of Technology, 02139, United States

E-mails: Mark.Polking@ll.mit.edu; liju@mit.edu

**Abstract:** Light sources and photodetectors operating in the far- to mid-infrared (FIR/MIR) band (8-12 μm, 0.1-0.15 eV) remain relatively poorly developed compared to their counterparts operating in the visible and near-infrared ranges, despite extensive application potential for thermal imaging, standoff sensing, and other technologies. This is attributable in part to the lack of narrow-gap materials (<0.1eV) with high optical gain and absorption. In this work, a narrow-gap semiconductor, $Pb_{0.7}Sn_{0.3}Se$, is demonstrated to exhibit an optical response >10× larger than that of $Hg_xCd_{1-x}Te$ (MCT), the dominant material for FIR/MIR photodetectors. A previous theoretical investigation indicated that chalcogen $p$ and metal $d$ band inversion in this material creates a Mexican hat band structure (MHBS), which results in a dramatic increase in the joint density of states at the optical transition edge compared to typical semiconductors. This prediction is experimentally validated here using single-crystal specimens of $Pb_{0.7}Sn_{0.3}Se$ measured using temperature-dependent spectroscopic ellipsometry over a wavelength range of 1.7-20 μm (0.73-0.062 eV). These measurements demonstrate a large enhancement in extinction coefficient and refractive index characteristic of a MHBS in the vicinity of the absorption edge, in agreement with theoretical predictions. The realization of topological semiconductors with a MHBS is expected to lead to high-efficiency detectors operating in the FIR/MID range.

The far- to mid-infrared (FIR/MIR) portion of the electromagnetic spectrum spanning the 8-12 micron (0.1-0.15 eV) wavelength band is of increasing importance for thermal imaging[1], chemical sensing[2–4], hyperspectral imaging[5], and numerous other critical application areas.

Despite the expanding application space of the FIR/MIR, however, technological progress in this spectral range continues to be outpaced by the rapid advances in the visible, near-IR (NIR), and MIR bands, in part due to a dearth of high-efficiency sources and detectors with low fabrication cost and high fabrication yield[6,7]. Rapid progress in the visible-MIR portion of the spectrum has been facilitated by the widespread availability of efficient, low-cost photodiode detectors and diode laser sources, which can be fabricated at scale from highly mature III-V and Group IV semiconductors[8]. FIR/MIR systems, in contrast, have long relied on detectors based on the ternary semiconductor $Hg_xCd_{1-x}Te$ (MCT), which suffers from poor fabrication yield and poor optical absorption[9], and on quantum cascade laser (QCL) sources, which are limited by similarly high fabrication costs, low power output, and low efficiencies[6,7].

A promising solution to this paucity of suitable FIR/MIR source and detector technologies is the development of narrow-gap semiconductor systems with enhanced optical absorption/gain, high defect tolerance, and crystal structures amenable to monolithic integration on established III-V and Group IV semiconductor platforms. Recently, several of the authors predicted a large enhancement in optical absorption and gain near the band edge in narrow-gap topological semiconductors with strong band inversion, characterized by a reversal of the typical energy ordering of the metal $d$ states and the chalcogen $p$ states in some regions of the $k$-space[10]. These materials exhibit narrow, direct band gaps induced by spin-orbit coupling with a typical magnitude of under 0.1 eV as well as a so-called "Mexican hat" band structure, illustrated schematically in **Figure 1**. The band topology of these materials results in a large enhancement of the joint density of states (JDOS) at the band edge, leading to an enhancement of optical absorption/gain by 1-2 orders of magnitude compared to semiconductors with conventional parabolic bands, such as MCT. The direct band gaps and strong optical responses in these materials provide an ideal combination of properties for achieving improved performance of FIR/MIR optical devices.

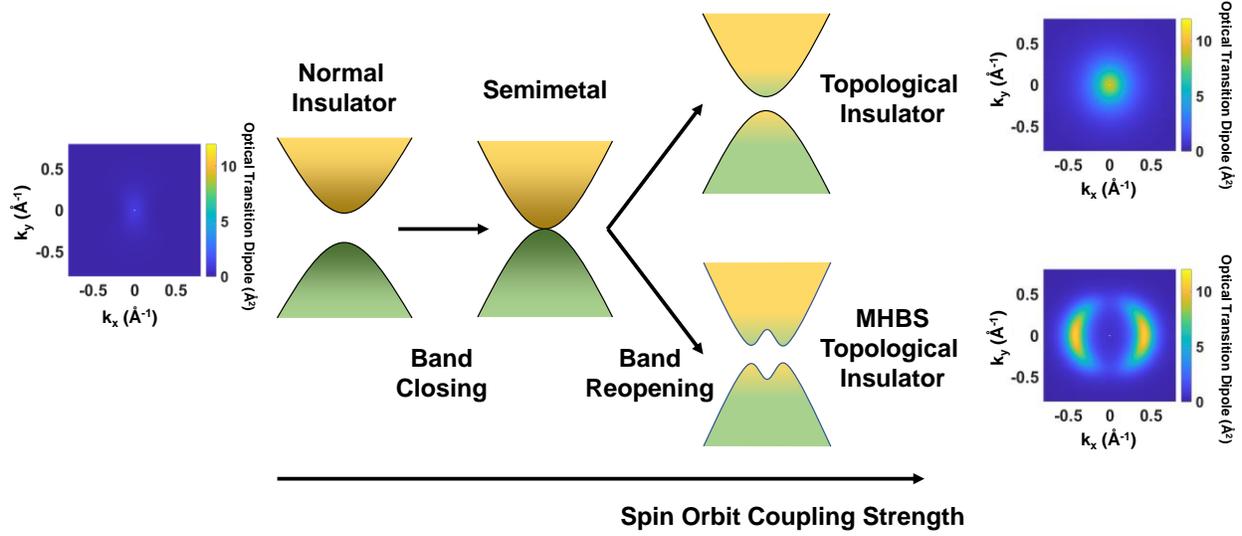

**Figure 1.** Emergence of a Mexican hat band structure (MHBS) with increasing spin-orbit coupling strength in a narrow-gap topological insulator (TI). The optical transition matrix element and joint density of staees are greatly enhanced in a TI with a MHBS (lower right) compared to topological semiconductors with parabolic band structure (upper right) and normal semiconductors (left). The optical transition dipole is defined as $|\langle ck|r|vk\rangle|^2$, where $r$ is the position operator, while $|ck\rangle$ and $|vk\rangle$ are the conduction and valence band wavefunctions, respectively. Adapted from Ref. [10].

In this work, we experimentally demonstrate strong optical absorption in an alloy of two of these predicted materials, SnSe and PbSe, stabilized in the rock-salt crystal structure predicted to yield a MHBS. This material system was chosen due to the ability to continuously tune the optical band gap over a wide range with material stoichiometry[11], and the ample heritage of lead salts in diode lasers systems and low-cost infrared detectors dating back to the 1960s[12]. High-quality single-crystal specimens of material with a composition of $Pb_{0.7}Sn_{0.3}Se$ were grown via the Bridgman method and characterized with temperature-dependent infrared spectroscopic ellipsometry. These measurements demonstrate an absorption coefficient as large as $2.0 \times 10^4$ cm$^{-1}$ at a wavelength of 12 μm, an enhancement of >10× compared to the band edge absorption of typical MCT samples with a CdTe mole fraction of 0.2 suitable for operation in the FIR/MIR[13]. Optical constants measured as functions of frequency and temperature with ellipsometry are in good agreement with predictions generated with linear response theory for varying values of carrier scattering lifetimes and prior predictions of Xu et al.[10] In addition to exceptional optical responses, this material also possesses a very large static dielectric constant (~200)[14]. High dielectric constants in similar lead chalcogenide materials have been shown to yield far superior defect tolerance compared with III-V materials and MCT[15], potentially enabling greatly improved device fabrication yield. In addition, the lattice constant of this material system enables direct monolithic heteroepitaxial integration on common III-V semiconductor substrates[16] and epitaxial integration on Si-Ge using alkaline earth halide buffer layers, as demonstrated for similar lead chalcogenides in prior reports[17,18]. This combination of narrow, compositionally tunable band gaps, exceptional responses, high defect tolerance, and compatibility with mature semiconductor platforms provide an ideal starting point for next-generation FIR/MIR sources and detectors.

As predicted previously by Xu *et al.*, both PbSe and SnSe in the rock salt phase exhibit a MHBS, suggesting that ternary alloys could provide a material system with a composition tunable band gap. Bulk SnSe, however, exists in an orthorhombic crystal structure derived from the higher symmetry rock salt structure by a small cell-doubling displacive transition.[19] A prior study indicates that SnSe-PbSe alloys must consist of at least 67% PbSe to achieve a single-phase material in the rock salt structure, which is stable for all compositions up to pure PbSe, and thus $Pb_{0.7}Sn_{0.3}Se$ was selected as a starting composition[20]. Single crystals of $Pb_{0.7}Sn_{0.3}Se$ were prepared from high-purity Sn, Se, and Pb feedstocks using a two-stage vertical Bridgman growth process. Further details are provided in the Supporting Information (SI), and images of typical crystal pieces are shown in **Figure S1**. Analysis of pulverized material with x-ray diffraction demonstrates single-phase material in the cubic rock salt structure, as illustrated in **Figure 2**a. Application of the extrapolation procedure of Nelson and Riley yields a cubic lattice constant of 6.0908 Å, reduced from the value of 6.12 Å for pure PbSe as expected.[21,22] Elemental analysis with energy-dispersive x-ray spectroscopy (EDS) in a scanning electron microscope (**Figure 2**b) demonstrates a composition of $(Pb_{0.74}Sb_{0.26})Se_{0.97}$, very close to the target composition within the typical error of EDS.

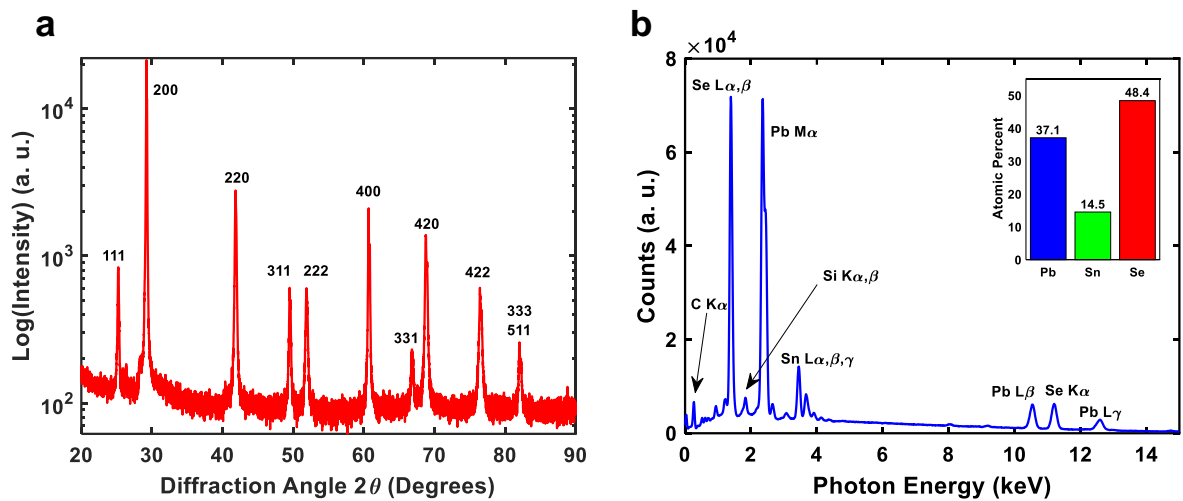

**Figure 2.** Structural and chemical characterization of a sample with a nominal composition of $Pb_{0.7}Sn_{0.3}Se$. a) X-ray diffraction scan of a pulverized single crystal of cubic-phase $Pb_{0.7}Sn_{0.3}Se$ grown by the Bridgman technique. All diffraction peaks can be assigned to the cubic rock salt structure expected for $Pb_{0.7}Sn_{0.3}Se$. b) Energy-dispersive x-ray spectroscopy (EDS) analysis of the stoichiometry of a $Pb_{0.7}Sn_{0.3}Se$ specimen. The measured composition matches the nominal target composition within the typical error of EDS.

Pieces of the bulk crystal were analyzed using infrared spectroscopic ellipsometry over a wavelength range of 1.7-20 μm (0.73-0.062 eV). Measurements were performed using a JA Woollam IR-VASE Mark II spectroscopic ellipsometer equipped with a broadband SiC globar source. The IR-VASE enables high-resolution measurements over a wavelength range of 1.7–30 μm, but the measurement range in this experiment was limited to 1.7-20 μm due to the limited transmission range of the ZnSe windows. The effects of the ZnSe windows on the ellipsometry measurement were accounted for using a JA Woollam proprietary calibration method prior to measurement. Due to the construction of the cryostat, the angle of incidence was fixed at 70°. Samples were scanned with a fixed polarizer angle of 45 degrees and a resolution of 16 cm⁻¹. Prior to measurement, the crystal pieces were mechanically polished along (100) planes and mounted on roughened Si substrates to suppress depolarization effects due to diffuse surface reflections and back surface reflections, respectively (**Figure S2**). The extinction coefficient and refractive index of $Pb_{0.7}Sn_{0.3}Se$ samples were measured in a cryostat under high vacuum at temperatures from 77 K up to ambient temperature (**Figure 3**). For the purposes of calculating *n* and *k* from the ellipsometric measurement data, the sample was treated as a single, homogenous layer.

Measurements of the extinction coefficient demonstrate a marked enhancement from 270 to 77 K in the vicinity of the band edge with a peak value of approximately 1.75 at an energy of 0.1 eV. The sharp peak observed is consistent with the predictions of Xu *et al.* for rock salt SnSe with a MHBS[10]. This peak is a characteristic of MHBS as conventional semiconductors with parabolic band structure should exhibit $\kappa$ that increase monotonically but slowly near the band edge. Measurements of the real part of the refractive index also show a large enhancement from 270 to 77 K, reaching a peak value of 6.5 at the latter temperature. Particulalry, the extinction coefficient $\kappa$ is larger than that of MCT by more than 10 times in the same frequency range (inst of Figure 3a) The measurements show low levels of depolarization (**Figure S3**), indicating low levels of diffuse surface scattering and back-surface reflections. We also calculated the peak value of the imaginary part of the dielectric constant $\varepsilon^{(2)}$ of $Pb_{0.7}Sn_{0.3}Se$ from the measured values for $n$ and $k$, which yields $\varepsilon^{(2)} \approx 22$. This is far larger than values predicted by Xu *et al.* for SnSe under 8% tensil strain, which is topologically trivial without band inversion (typically $\varepsilon^{(2)} < 3$ near the band edge). Meanwhile, the measured $\varepsilon^{(2)}$ of $Pb_{0.7}Sn_{0.3}Se$ is still smaller than the predicted value for unstrained SnSe by Xu *et al.* (roughly 60), which is topologically nontrivial with MBHS. This may be attributable to alloying with PbSe, which is predicted to exhibit an optical gain that is ~35% smaller than the value predicted for pure SnSe ($2.6 \times 10^4$ cm$^{-1}$ vs. $4.0 \times 10^4$ cm$^{-1}$) in the rock-salt phase. Also, the calculations by Xu *et al.* assumed zero temperature and an electron relaxation time of 0.8 ps. At finite temperature with potentially shorter electron relaxation time, which is relevant to our experiments, the peak of $\varepsilon^{(2)}$ would be lower (cf. Figure 4b).

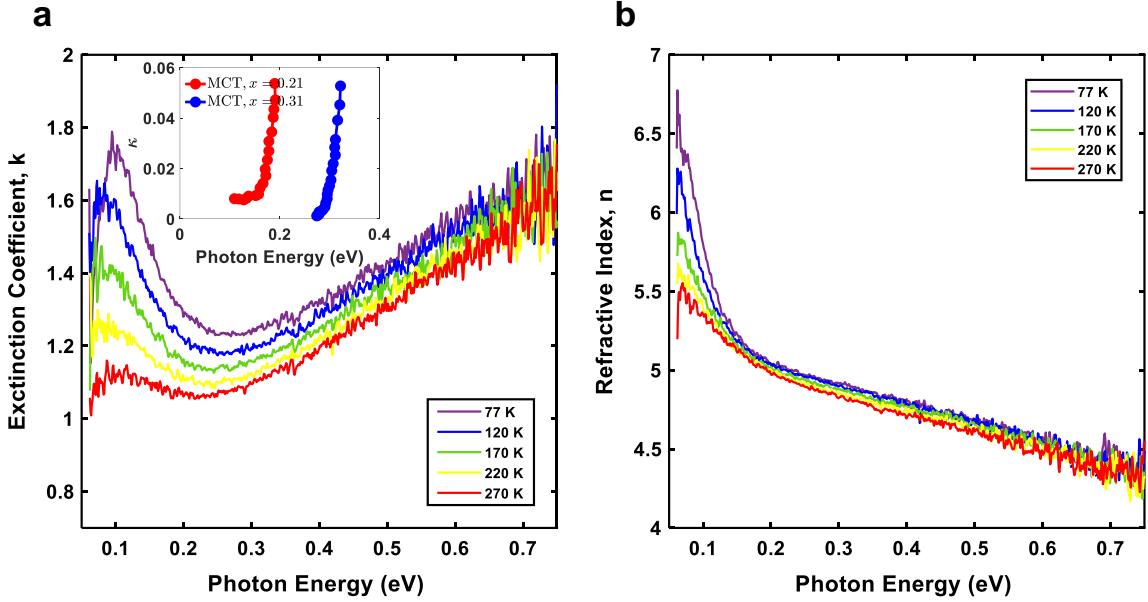

**Figure 3.** Temperature-dependent spectroscopic ellipsometry measurements of a rock-salt $Pb_{0.7}Sn_{0.3}Se$ single crystal specimen from 77 K to 270 K illustrating an optical response characteristic of a Mexican hat band structure (MHBS). a) Extinction coefficient as a function of photon energy, and b) real part of the refractive index as a function of photon energy. Large enhancements of both the real and imaginary parts of the refractive index can be observed near the band edge. Particularly, there is a sharp peak in $\kappa$ near the band edge. This should be compared with semiconductors with conventional parabolic band structure, where $\kappa$ would increase monotonically but slowly near the band edge. This is consistent with prior predictions of the optical response of a topological semiconductor with a MHBS[10] (cf. Figure 4a), and is a characteristic of MHBS. Inst of (a) shows the extinction coefficient of $Hg_xCd_{1-x}Te$ (MCT), which are from Ref. [23]

The temperature dependences of the optical constants observed here are inconsistent with thermorefractive effects common to semiconductors. A large enhancement of the refractive index is observed only within ~0.1 eV of the band edge, suggesting that thermo-refractive effects common to semiconductors are not responsible for the observed behavior. The increase in refractive index of ~0.1 observed at higher energies upon cooling from 270 to 77 K is consistent with prior measurements of pure PbSe, which exhibits negative thermorefraction[18]. An increase in free carrier absorption can also be excluded as a source of this temperature dependence. Increasing free carrier absorption results in an increase in the extinction coefficient, but a coincident decrease in the real part of the refractive index, in stark contrast to our results showing marked increases in both quantities with decreasing temperature. The sharp absorption peak observed is also inconsistent with the typical $1/f^2$ dependence of free carrier absorption according to the Drude model[24].

To better understand the experimental results and elucidate the origin of the temperature dependence, we performed first-principles calculations of the linear optical responses (details in SI). Due to the high computation cost for simulations of $Pb_{0.7}Sn_{0.3}Se$ alloys, we used SnSe as a test case to illustrate the behavior of the optical response of a MHBS topological insulator near the band-edge and its temperature dependence. Specifically, we first calculated the dielectric response function using the linear response theory, which can be expressed as

$$\varepsilon_{pq}(\omega) = \delta_{pq} - \frac{e^2}{\varepsilon_0} \int \frac{d^3\boldsymbol{k}}{(2\pi)^3} \sum_{m,n} \frac{(f_m - f_n)\langle m|r^p|n\rangle\langle n|r^q|m\rangle}{E_n - E_m - \hbar\omega - i\eta}, \tag{1}$$

where $r^{p,q}$ is the position operator with Cartesian indices $p$ and $q$. $\delta_{pq}$ is the Kronecker delta, $\varepsilon_0$ is the vacuum permittivity and $f_m$ and $E_m$ are the occupation number and energy of the electronic state $|m\rangle$, respectively. The dependence on electron wavevector $\boldsymbol{k}$ is omitted for simplicity. The parameter $\eta = \frac{1}{\tau}$ is the linewidth of the electrons, and $\tau$ is the carrier relaxation time. Here we assume $\tau$ is a constant for all electrons (constant relaxation time approximation). Then, the extinction coefficient $k$ and refractive index $n$ can be obtained from the equation:

$$n(\omega) + ik(\omega) = \sqrt{\varepsilon(\omega)} \tag{2}$$

It should be noted that in a perfect crystal, changes in temperature primarily result in two consequences. First, the occupation number $f_m$ follows the Fermi-Dirac distribution. The bandgap of SnSe ($Pb_{0.7}Sn_{0.3}Se$) is around 0.1 eV, equivalent to over 1000 K. Hence, the temperature variance between 77 K and 270 K does not lead to a significant change in $f_m$. On the other hand, temperature variations can also affect the carrier scattering time $\tau$, which is typically substantially reduced at elevated temperatures[25,26]. Since it is not straightforward to predict $\tau$ as a function of temperature, especially for alloy systems, we manually vary $\tau$ and calculate $n(\omega)$ and $k(\omega)$ for different $\tau$ values. The results are shown in **Figure 4**. Similar to the experimental results, peaks in both $n(\omega)$ and $k(\omega)$ are clearly observable when $\omega$ is close to the band-edge (around 0.2 eV for SnSe in the rock salt phase in our calculations). These peaks come from the fast electronic interband transitions near the band-edge and are a characteristic feature of topological semiconductors with a MHBS, which leads to a large joint density of state. As shown in our simulations, when $\tau$ becomes shorter, corresponding to a higher temperature, the peaks in $n$ and $k$ become less prominent, consistent with experimental results as well. This reduction in peak magnitude for shorter values of $\tau$ (larger $\eta$) is due to the smearing of the electronic states near the band edge.

Our results demonstrate the feasibility of achieving a large enhancement of optical responses in the FIR/MIR spectral region due to ample $d$-$p$ orbital character mixing arising from band inversion in topological insulators (TI), and a special Mexican hat band structure in the rock-salt phase of PbSe-SnSe alloys, consistent with prior theoretical studies. The direct band gap in this class of TI is proportional to the spin-orbit coupling strength, which has a suitable magnitude for FIR/MIR devices. Our measurements of the optical constants of high-quality single crystals of the topological semiconductor $Pb_{0.7}Sn_{0.3}Se$ with spectroscopic ellipsometry clearly demonstrate a pronounced absorption peak near the band edge and coincident enhancement of the refractive index, arising from (a) an inordinately large $d$-$p$ mixing due to the topological band inversion across the direct band gap, which enhances the optical transition matrix element, and (b) the high JDOS of a MHBS material, for which the optical transition can occur on a ring of degenerate states rather than at a single $\mathbf{k}$-point, as for a standard TI without a MHBS (**Figure 1**). The temperature dependence of the optical constants closely matches the results of first-principles calculations of the temperature-dependent optical response of a MHBS semiconductor with free carriers derived from linear response theory.

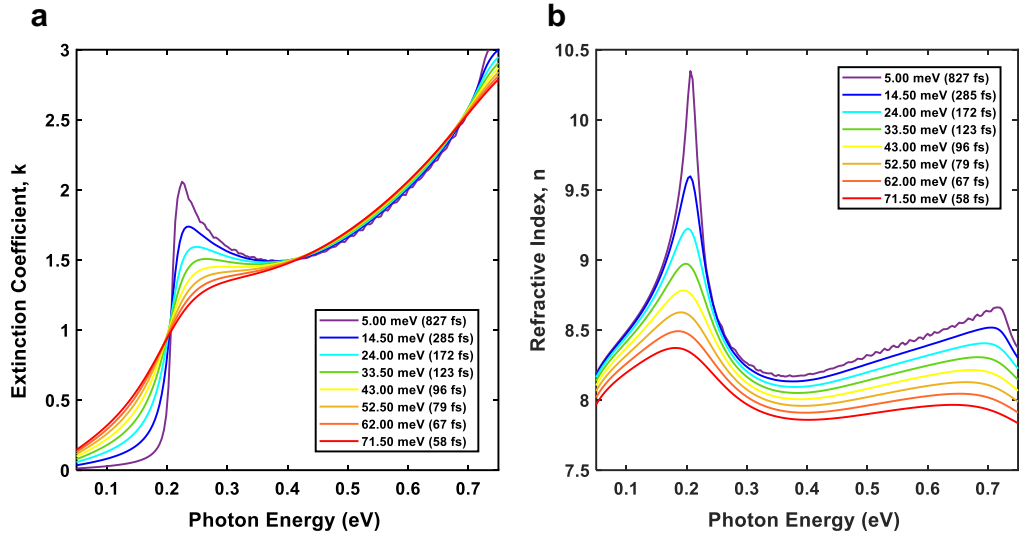

**Figure 4.** Simulated optical constants of SnSe in the rock salt structure for varying carrier scattering times. a) Extinction coefficient vs. photon energy, and b) the real part of the refractive index vs. photon energy. The predicted behavior of both the real and imaginary parts of the refractive index is in close agreement with the experimental measurements shown in **Figure 3**.

This work highlights a pathway to the engineering of large photonic responses in materials with band gaps suitable for the underdeveloped FIR/MIR spectral range. As shown previously[10], numerous other material systems are predicted to exhibit a similar MHBS associated with strong optical responses. Further development of these materials is anticipated to enable next-generation light sources and photodetectors with improved optical gain and detection efficiency, respectively, with broad applications to thermal imaging, standoff chemical sensing, hyperspectral imagers, and beyond.

**Acknowledgments:**

M.J.P acknowledges Michael Walsh for assistance with optical polishing of specimens. R. S. acknowledges the financial support provided by the Ministry of Science and Technology in Taiwan under Project No. NSTC-113-2124-M-001-003 and No. NSTC-113-2112M001-045-MY3, as well as support from Academia Sinica for the budget of AS-iMATE11412. J.L. and H-W.X. acknowledge support by the Office of Naval Research Multidisciplinary University Research

**Conflicts of Interest:**

The authors declare no conflicts of interest.

**Data Availability Statement:**

The data that support the findings of this study are available from the corresponding authors upon reasonable request.

**Keywords:**

Optical materials, infrared detectors, long-wave infrared, ellipsometry, lead salts, Bridgman growth